\documentclass{ccjnl}

\title{Multi-User Semantic Fusion for Semantic Communications over Degraded Broadcast Channels}
\author{Tong Wu\inst{1}, Zhiyong Chen\inst{1,*}, Meixia Tao\inst{1,*}, Bin Xia\inst{1}, Wenjun Zhang\inst{1}\corinfo{\{zhiyongchen, mxtao\}$@$sjtu.edu.cn}}
\receiveddate{}
\reviseddate{}
\Editor{}
\address[1]{Cooperative Medianet Innovation Center, Shanghai Jiao Tong University, Shanghai, China}

\begin{document}
\maketitle

\begin{abstract}
Degraded broadcast channels (DBC) are a typical multiuser communication scenario, Semantic communications over DBC still lack in-depth research. In this paper, we design a semantic communications approach based on multi-user semantic fusion for wireless image transmission over DBC. In the proposed method, the transmitter extracts semantic features for two users separately. It then effectively fuses these semantic features for broadcasting by leveraging semantic similarity. Unlike traditional allocation of time, power, or bandwidth, the semantic fusion scheme can dynamically control the weight of the semantic features of the two users to balance the performance between the two users. Considering the different channel state information (CSI) of both users over DBC, a DBC-Aware method is developed that embeds the CSI of both users into the joint source-channel coding encoder and fusion module to adapt to the channel. Experimental results show that the proposed system outperforms the traditional broadcasting schemes.
\end{abstract}

\section{INTRODUCTION}

In recent years, semantic communications have received significant attention from both industry and academia. With the help of artificial intelligence (AI), semantic communications can extract the semantic information from the original data, and further transmit it, thereby significantly improving communication efficiency. Therefore, semantic communications have been considered a promising solution for the sixth-generation (6G) wireless networks \cite{ZPing,shen2}.

Several studies have been conducted on semantic communications for different types of original information, such as text \cite{text2021, text2022}, image \cite{gundu2019,image2022, image2022tao,Yang,KeYang,Dai2,CDDM}, and video \cite{gundu2022video, video2022, video2022-2}. For text transmission, a deep learning-based semantic communication system is proposed in \cite{text2021}, named DeepSC, which has an advantage in the low signal-to-noise ratio (SNR) regime. For image transmission, \cite{gundu2019} first introduces a CNN-based deep joint source-channel coding (JSCC) scheme that exhibits superior performance compared to separation-based digital transmission schemes. In \cite{ADJSCC}, the ADJSCC model, which incorporates channel state information (CSI) with a channel-wise soft attention scheme, is introduced. \cite{KeYang} reengineers the JSCC architecture incorporating with advanced Swin Transformer\cite{swinTransformer}, enhancing the model capacity. This paper also designs a deep channel attention module. It utilizes several attention operations to dynamically scale the semantic features based on CSI. This achieves stable reconstruction quality under varying SNRs. However, these channel adaptive methods based on attention introduce significant additional computational burden and parameters to the overall system. This is due to the high computational complexity of attention operations, as well as their application in high-dimensional feature spaces. For video transmission, the end-to-end JSCC video transmission scheme is proposed in \cite{gundu2022video}. Then, \cite{video2022} designs a novel deep joint source-channel coding approach to achieve wireless video transmission, which can outperform traditional wireless video coded transmission schemes.


It is worth noting that previous works mainly focus on point-to-point semantic communications, while research on multi-user semantic communications is relatively limited. For multiple access channel, a novel multiple access technology termed model division multiple access (MDMA) is proposed in \cite{MDMA}. By leveraging resources in the semantic domain, MDMA can achieve greater performance gains compared to traditional multiple access methods.
For broadcasting channels, a one-to-many scheme is proposed in \cite{Hu} for text transmission, where the transmitter concatenates these texts together and extracts their semantic features for transmission. For relay channels, a semantic-and-forward scheme, as designed in \cite{relay}, assists the relay node in forwarding semantic information and addresses the differences in background knowledge between the source and destination nodes. Then, a novel deep joint source-channel coding scheme for image transmission over a half-duplex cooperative relay channel is presented in \cite{gundu2022relay}.

Actually, multiuser semantic communications are not simply point-to-point semantic communications but require corresponding design for effective multi-user transmission scheme and channel adaptive method. Motivated by this, we consider a degraded broadcast channel (DBC) in this paper, which is a typical multiuser communication scenario. There is a transmitter and multiple users located in different geographical locations. There are numerous traditional transmission schemes to divide the channel between two users, such as time division (TD), frequency division (FD) and superposition coding with {successive interference cancellation (SIC)} which is a capacity region achieving scheme. However, in-depth research remains lacking regarding semantic communication scheme over DBC and multi-user channel adaptive method.

To address this issue, we propose a fusion-based semantic communications system over two-user DBC with a channel adaptive method called DBC-Aware for wireless image transmission. In the proposed architecture, considering the similarity in the semantic features of images required by both users, we develop a multi-user semantic fusion (SF) scheme based on Transformer. Moreover, considering the different CSI of both users over DBC, we design a DBC-Aware method that embeds the CSI of both users into the JSCC encoder and fusion module to adapt to the varying channel conditions.

The contributions of this paper can be summarized as follows.
\begin{figure*}[t]
\centering
\includegraphics[width=1\textwidth]{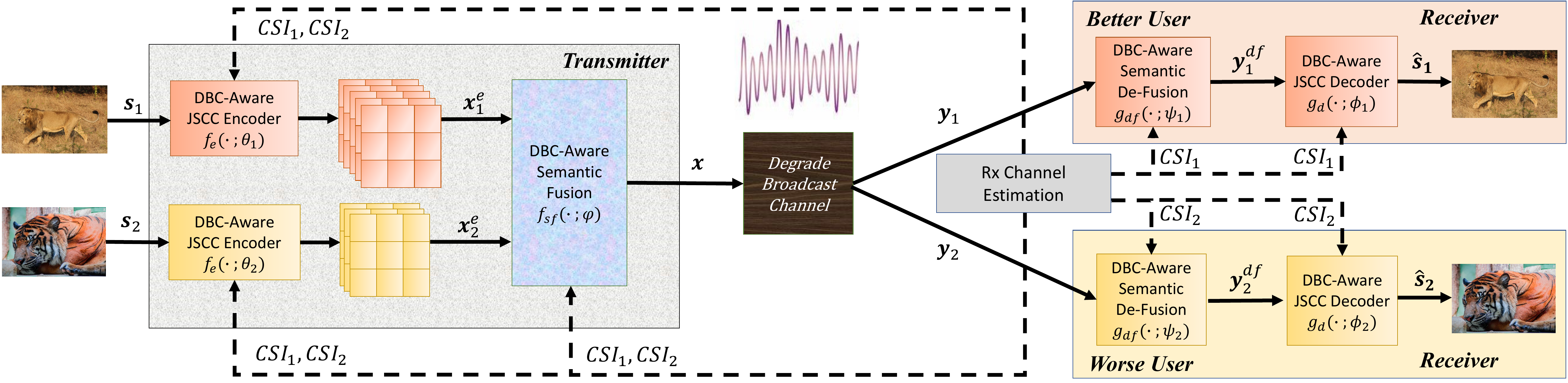}
\caption{The overall architecture of the proposed fusion-based multi-user semantic communication system over DBC}
\label{model_structure}
\end{figure*}

\begin{itemize}
    \item \textbf{Transmission framework:} We propose a fusion-based semantic communications system over DBC for wireless image transmission. In the proposed architecture, a transmitter can extract and fuse the semantic features of both users as a joint latent representation, incorporating with the CSI of DBC for broadcasting. Both users can de-fuse the joint representation and then decode their individual images with their own CSI.
    \item \textbf{Multi-user semantic fusion:} We investigate the semantic similarity between semantic features from both users measured by $RV_2$-coefficient \cite{RV2} and the Pearson correlation coefficient. Accordingly, we develop the DBC-Aware semantic fusion module based on Transformer, consisting of three stages to facilitate the deep fusion of multi-user semantic. Unlike traditional resource allocation methods such as time, power, or bandwidth, the semantic fusion scheme can dynamically control the weight of the two users' semantic features in the joint latent representation to balance the performance between the two users.

    \item \textbf{DBC-Aware channel adaptive method:} We design a novel multi-user channel adaptive method to align each user's semantic features with their respective channels, accommodating different SNRs. By embedding CSI into encoder and decoder in the transmitter and receiver, respectively, we achieve remarkable performance across a range of SNRs with a single model. Furthermore, the proposed method only increases few parameters and computational complexity, which is achieved by leveraging computationally efficient additive operations, reducing the coding dimensions of CSI, and reusing modules.

  \end{itemize}
The rest of this paper is organized as follows. The system model is introduced in Section II, including the architecture of the DBC-Aware JSCC. The fusion based broadcasting scheme for semantic communications is proposed in Section III. Extensive experimental results are presented in Section IV. Finally, {conclusions and furture work are drawn in Section V}.

\section{SYSTEM MODEL}
\indent In this section, we propose a fusion-based multi-user semantic communications system for wireless image transmission over degraded broadcast channels.
The system involves one transmitter and two users, with the objective of delivering two distinct image messages  $\mathbf{s}_1$ and $\mathbf{s}_2$ to the respective users through semantic communications.
We establish a DBC-Aware JSCC, achieving channel adaptability to DBC.
\subsection{System Overview}
For the transmitter, two DBC-Aware JSCC encoders, $f_e(\cdot;\theta_1)$ and $f_e(\cdot;\theta_2)$, extract semantic information from the source images $\mathbf{s}_1\in \mathbb{R}^{3 \times H \times W}$ and  $\mathbf{s}_2 \in \mathbb{R}^{3 \times H \times W}$, and encode it as semantic features $\mathbf{x}^e_1 \in \mathbb{R}^{c_1 \times h \times w}$ and $\mathbf{x}_2^e \in \mathbb{R}^{c_2 \times h \times w}$ respectively, as shown in Fig. \ref{model_structure}. Here, $c_1$ and $c_2$ represent the channel numbers, which are determined by the fusion ratio $\alpha$. The fusion ratio controls the reconstruction quality for two users. $h$ and $w$ denote the height and width of each channel, while $H$ and $W$ represent the height and width of the source images, with 3 indicating the number of color channels.
$f_e$ represents the model structure and $\theta_1,\theta_2$ encapsulate the network parameters. By leveraging the CSI of the DBC as inputs, the DBC-Aware JSCC encoders are designed to dynamically extract and encode semantic information in an adaptive manner, catering to the varying DBC states. Subsequently, we propose a DBC-Aware SF module $f_{sf}(\cdot;\varphi)$ to broadcast the information of $\mathbf{x}_1^e$ and $\mathbf{x}_2^e$ simultaneously by fusing them. By exploiting their semantic similarity along with the corresponding CSI, the module $f_{sf}(\cdot;\varphi)$ effectively and adaptively fuses the two semantic features into a joint latent representation $\mathbf{x} \in \mathbb{C}^n$, where $n$ signifies the channel use number and $\varphi$ represents network parameters. Then, the joint latent representation $\mathbf{x}$ is power-normalized and broadcasted to both users.

In this paper, we consider two distant users with different AWGN channels. The channel for user $1$ is an AWGN channel with noise power of $\sigma_1^2$, whose CSI denotes $CSI_1$. Similarly, the channel for user $2$ is also an AWGN channel with noise power $\sigma_2^2$, whose CSI is $CSI_2$. Without loss of generality, we assume that $\sigma_1^2 < \sigma_2^2$, such that we refer to user $1$ as the better user and user $2$ as the worse user. The received signals of the two users are $\mathbf{y}_1=\mathbf{x}+\mathbf{n}_1$ and $\mathbf{y}_2=\mathbf{x}+\mathbf{n}_2$ respectively, where $\mathbf{n}_1$ and $\mathbf{n}_2$ are Gaussian noise with noise power $\sigma_1^2$ and $\sigma_2^2$, respectively.

\begin{figure*}[t]
    \centering
    \includegraphics[width=1\textwidth]{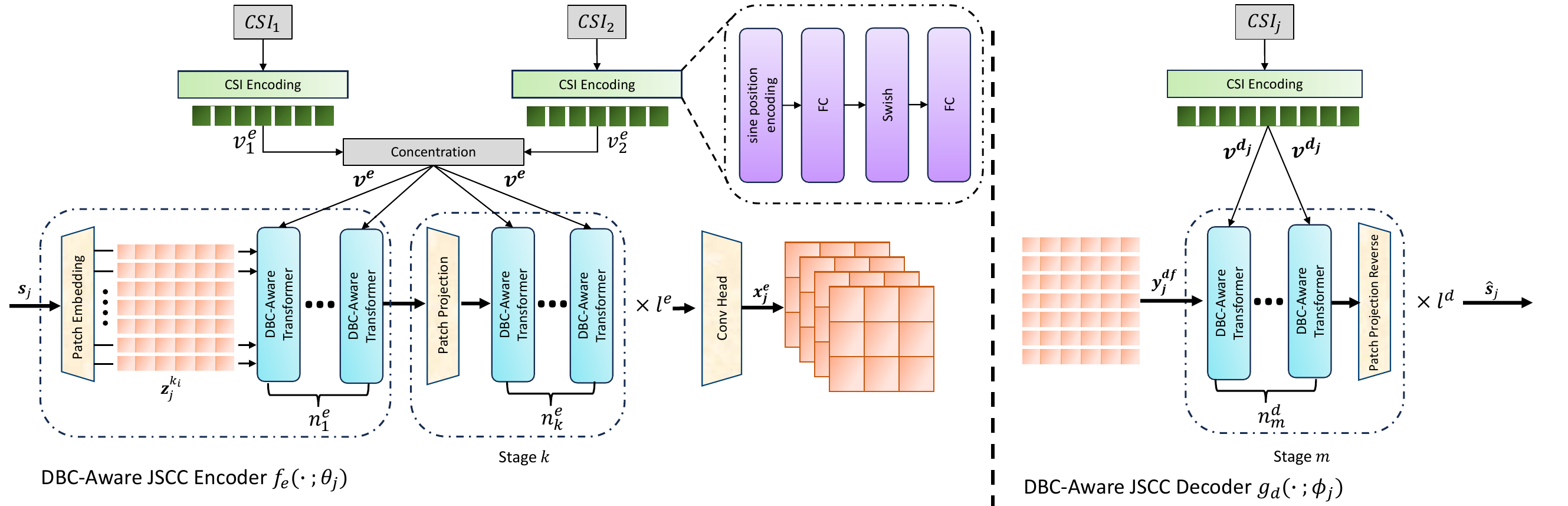}
    \caption{The overall structure of the proposed DBC-Aware JSCC}
    \label{JSCC}
\end{figure*}
\begin{figure}[t]
    \centering
    \includegraphics[width=0.48\textwidth]{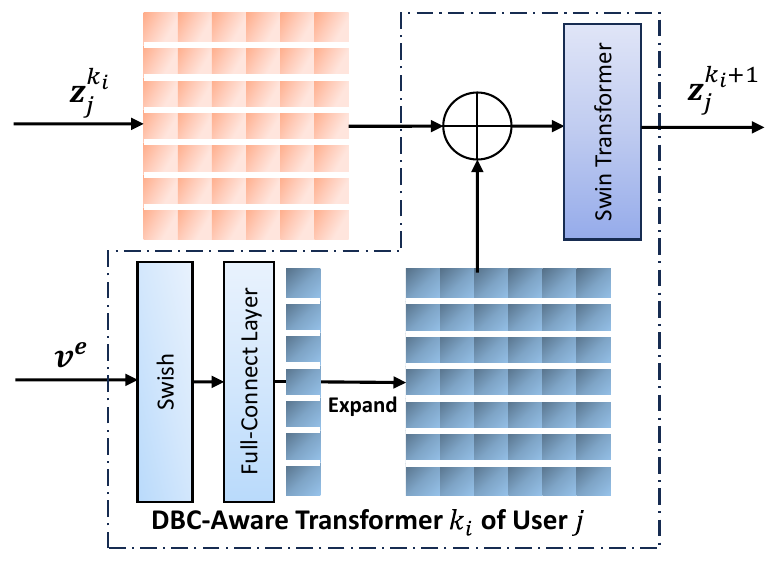}
    \caption{The architecture of DBC-Aware Transformer}
    \label{DBC_Aware_Transformer}
\end{figure}

The better user first processes the received signal $\mathbf{y}_1$ as $\mathbf{y}_1^{df}$ utilizing a DBC-Aware Semantic De-Fusion module $g_{df}(\cdot;\psi _1)$. Then, a DBC-Aware JSCC decoder $g_d(\cdot;\phi _1)$ is used to reconstruct the source image $\mathbf{s}_1$ as $\hat{\mathbf{s}}_1 \in \mathbb{R}^{3 \times H \times W}$.
Converserly, the worse user attempts to de-fuse its received signal $\mathbf{y}_2$ as $\mathbf{y}_2^{df}$ using the module $g_{df}(\cdot;\psi _2)$. Following the de-fusion step, $\mathbf{y}_2^{df}$ is decoded with $g_d(\cdot;\phi_2)$ to reconstruct the source image $\mathbf{s}_2$ as $\hat{\mathbf{s}}_2 \in \mathbb{R}^{3 \times H \times W}$. Here, $\psi_1, \psi_2$ and $\phi_1,\phi_2$ are network parameters.  Meanwhile, each user is only able to estimate their own CSI knowledge. Consequently the better user is limited to de-fusing and decoding the signal exclusively with $CSI_1$ while the worse user undergoes both processes utilizing $CSI_2$.

In this paper, neural networks are employed for all modules, and the structural design plays a crucial role in enhancing the performance of the semantic communication system over DBC. In the following, we elaborate on the design of the DBC-Aware JSCC.
\subsection{The Architecture of the DBC-Aware JSCC}

To enhance the performance of JSCC for both users under different channel conditions, we propose the DBC-Aware JSCC, which utilizes the proposed DBC-Aware method to integrates CSI succinctly and efficiently into the JSCC, achieving channel adaptability without adding excessive parameters and computational overhead. Fig. \ref{JSCC} illustrates the overall architecture of the proposed DBC-Aware JSCC encoder $f_e(\cdot;\theta_j)$ and decoder $g_d(\cdot;\phi_j)$ for $j=1,2$. 

The DBC-Aware JSCC encoder involves two CSI encoding modules and $l^e$ stages. 
The two CSI encoding modules encode $CSI_1$ and $CSI_2$ into two CSI vectors $\mathbf{v}^e_1,\mathbf{v}_2^e \in \mathbb{R}^{t}$, respectively. These vectors are concatenated as $\mathbf{v}^e \in \mathbb{R}^{2t}$ and then fed into all the DBC-Aware Transformers in the JSCC encoder. On the other hand, the source image $\mathbf{s}_j$ is fed into stage 1, including a patch embedding block which transforms it into a feature map with size ${h_{1}^e \times w_{1}^e}$ and $n_1^e$ DBC-Aware Transformers applied on the feature maps, incorporating the CSI vector $\mathbf{v}^e$. 
The transformer blocks maintain the size of the input feature maps. The output feature map of stage $1$ is fed into stage $2$, where a patch projection block resizes the feature map into ${h_{2}^e \times w_{2}^e}$ and then $n_2^e$ DBC-Aware Transformer blocks with additional input $\mathbf{v}^e$ is
applied afterwards for feature transformation. The stage is repeated for $(l^e-1)$ times, as stage $k,k=2,3,...,l^e$, with $n_k^e$ transformer blocks outputting feature map $\mathbf{z}_j^{k_i} \in \mathbb{R}^{h_k^e \times w_k^e}$.
Finally, a CNN layer is applied on the feature map, resulting in the output $\mathbf{x}_j^e$.

At the user $j$, user $j$ only needs to input their own $CSI_j$ to the CSI encoding module, resulting in $\mathbf{v}^{d_j} \in \mathbb{R}^{u}$. The DBC-Aware JSCC decoder takes the feature map $\mathbf{y}_j^{df}$ as input, applying $n_1^d$ DBC-Aware Transformers that incorporate $\mathbf{v}^{d_j}$. 
Subsequently, the reverse projection block resizes the feature map into $h_1^d \times w_1^d$. The $n_1^d$ transformer blocks together with the reverse projection block are refered as stage 1. This procedure is repeat $l^d$ times, denoted as stage $m,m=1,2,..,l^d$, with $n_m^d$ transformer blocks and output size of $h^d_m \times w^d_m$. It is noticable that the output feature map of stage $l^d$ has $h^d_{l^d} \cdot w^d_{l^d}$ elements, which is equal to $3HW$, meaning that this feature map can be reshaped as $\hat{\mathbf{s}}_j$ for reconstruction.

The core components of the DBC-Aware channel adaptive method include the novel CSI encoding modules and the DBC-Aware Transformer. The CSI encoding block begins by transforming the input $CSI_j$ into a vector representation by utilizing sine position embedding. Then, a full-connection (FC) layer, a swish activate function \cite{swish} and another FC layer are applied on the vector representation sequentially, outputting CSI vector $\mathbf{v}_j^e$ at the transmitter end and $\mathbf{v}^{d_j}$ at user $j$. The Swish activate function can be formulated as
\begin{equation}
    swish(x)=x \cdot sigmoid(x).
\end{equation}

The architecture of the DBC-Aware Transformer is depicted in Fig. \ref{DBC_Aware_Transformer}. For the $i$-th DBC-Aware Transformer at stage $k$ within the JSCC encoder, the input feature map dimensions are $h^e_k \times w^e_k$. The CSI vector $\mathbf{v}^e$ is transformed into a new vector with dimensions $h^e_k \times 1$ by a swish function and an FC layer for fusing the two CSI vectors $\mathbf{v}_1^e$ and $\mathbf{v}_2^e$. Following that, $w^e_k$ copy is generated to form the CSI map with the same size as the input feature map. The two maps are added together in element wise and fed into the backbone network. In this paper, we adopt the swin transformer as the backbone for its enhancing capacity to capture global dependencies, which has been prove to improve the performance for semantic communications. The transformers in the JSCC decoder of user $j$ have the same architecture with the transformers in JSCC encoder, except that the input CSI vector $\mathbf{v}_{d_j}$ is only computed from $CSI_j$. 

\begin{figure}[t]
    \centering
    \includegraphics[width=0.48\textwidth]{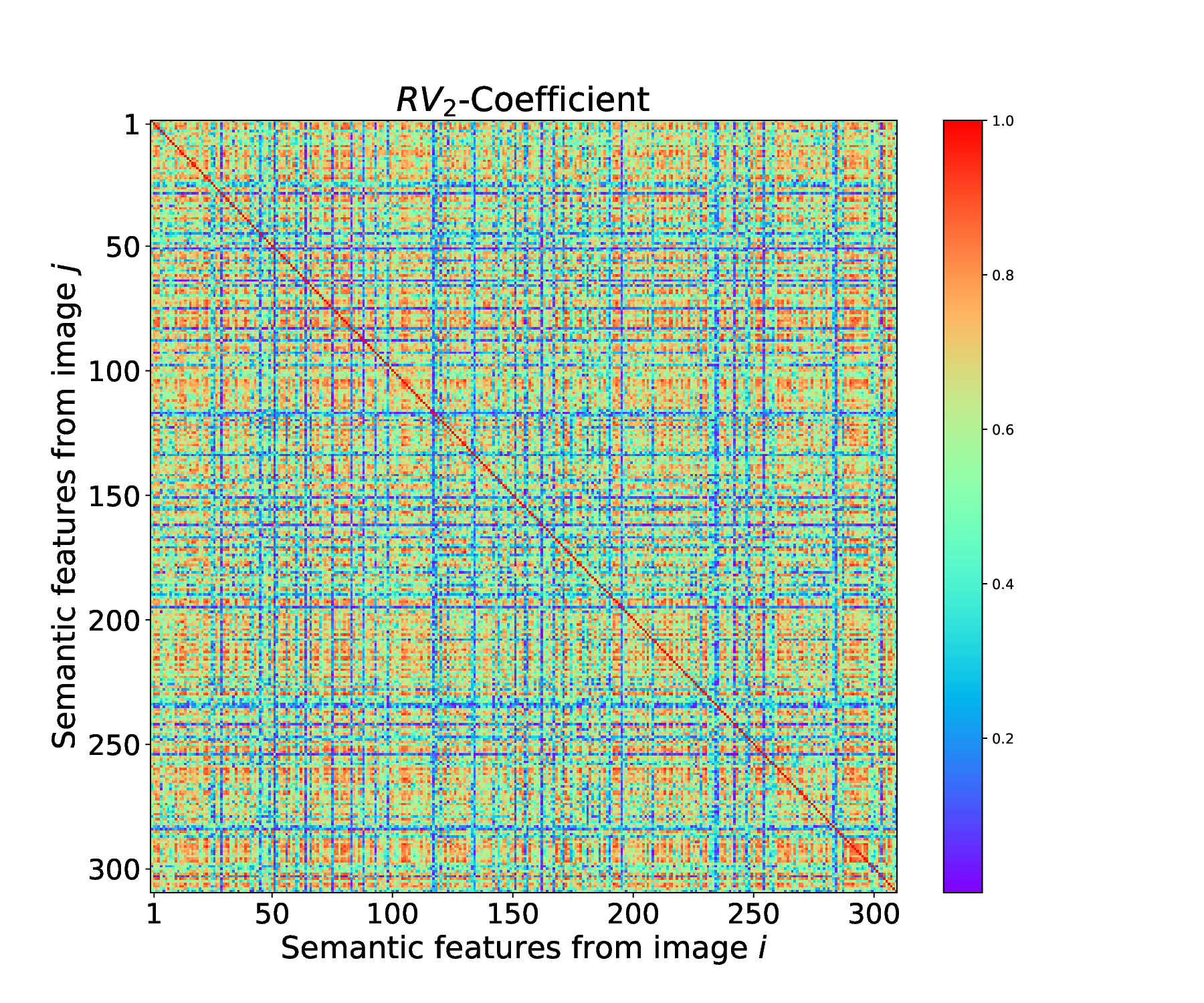}
    \caption{The $RV_2$-coefficient of semantic features from images in CelebA dataset}
    \label{RV2_coeff}
\end{figure}
\begin{figure}[t]
    \centering
    \includegraphics[width=0.48\textwidth]{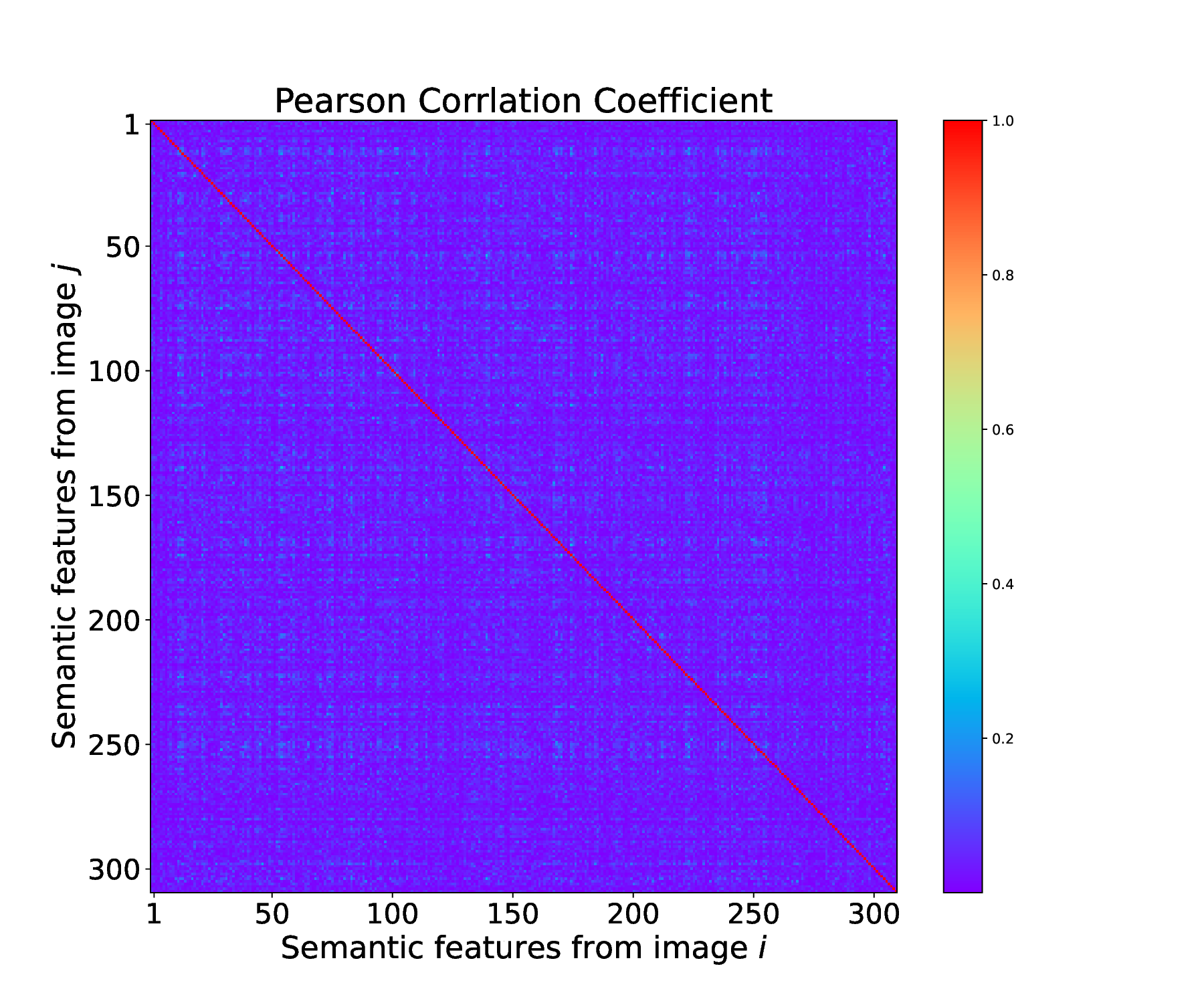}
    \caption{The Pearson correlation coefficient of semantic features from images in CelebA dataset}
    \label{Linear_coeff}
\end{figure}

The proposed DBC-Aware channel adaptive method seamlessly integrates CSI to achieve channel adaptation to DBC. This integration involves infusing CSI into each transformer at every stage. Additionally, each transformer is equipped with a module featuring independent parameters to effectively learn the utilization of CSI. 
{Encoding CSI as high-dimensional vectors is aimed at aligning with the model structure, thus ensuring the effective utilization of CSI.}
This approach ensures that all transformers can precisely acquire CSI, thereby achieving a comprehensive fusion of CSI for the extraction and encoding of semantic features in an efficient manner. Moreover, the preceding layer is dedicated to extracting and encoding semantic features exclusively, eliminating the need to propagate CSI to the subsequent layer. This strategy conserves model capacity and ensures performance. Because the data size of CSI is significantly smaller than that of an image, the entire model utilizes only two shared CSI encoding modules to inject CSI into all transformers. 
Furthermore, within each transformer, this information is mapped to a space with substantially lower dimensionality compared to the feature space and integrated with the semantic features by low-complexity additive operations. Therefore, the DBC-Aware channel adaptive method introduces only a little number of extra parameters and computational overhead.

\section{FUSION BASED BROADCASTING SCHEME FOR SEMANTIC COMMUNICATIONS}
In this section,we provide a detailed exposition of the introduced SF broadcasting scheme
and furtherly derive a tailored loss function according to the distinctive attributes of the DBC channel, to effectively train the entire network.

\subsection{Semantic Fusion Broadcasting Scheme}

When two users share the same channel via a semantic communications system, the semantic performance of the system can no longer be characterized by single number. Inspired by rate region, a semantic performance region is proposed to characterize the system, where each point in the region is a vector of achievable semantic performance that can be maintained by the both users simultaneously.

To enlarge the semantic performance region, the design of the broadcasting scheme necessitates the joint consideration of both source images and CSI. For both image sources $\mathbf{s}_1$ and $\mathbf{s}_2$, experiments reveal a semantic similarity between $\mathbf{s}_1$ and $\mathbf{s}_2$. As shown in Fig. \ref{RV2_coeff} and Fig. \ref{Linear_coeff}, we extract several hundred images from the CelebA dataset and employ the same DBC-Aware JSCC encoder to extract and encode their semantic features under identical CSI. As illustrated in Fig. \ref{JSCC}, the encoding results take the form of two-dimensional matrices. We compute the $RV_2$-coefficient and Pearson correlation coefficient between these matrices, showing the absolute values since we are only interested in the magnitude of correlation. We can find that most regions in the Fig. \ref{RV2_coeff} exhibit shades of green or red-orange, indicating a high $RV_2$-coefficient among the semantic features of these images. Only a few blue lines are present, signifying a low $RV_2$-coefficient of the semantic features of this image with all other images. Consequently, we can conclude that, overall, there exists semantic similarity among the images. However, the extensive blue areas in Fig. \ref{Linear_coeff} demonstrate a low linear correlation among the semantic aspects of the images. Hence, a straightforward summation for the fusion of their semantics is not suitable. It necessitates the design of a neural network to achieve efficient integration of semantic features.

Based on this observation, we design the SF broadcasting scheme, which utilizes the proposed DBC-Aware SF module to broadcast the semantic information of both users in a fusion manner.
\begin{figure}[t]
    \centering
    \includegraphics[width=0.48\textwidth]{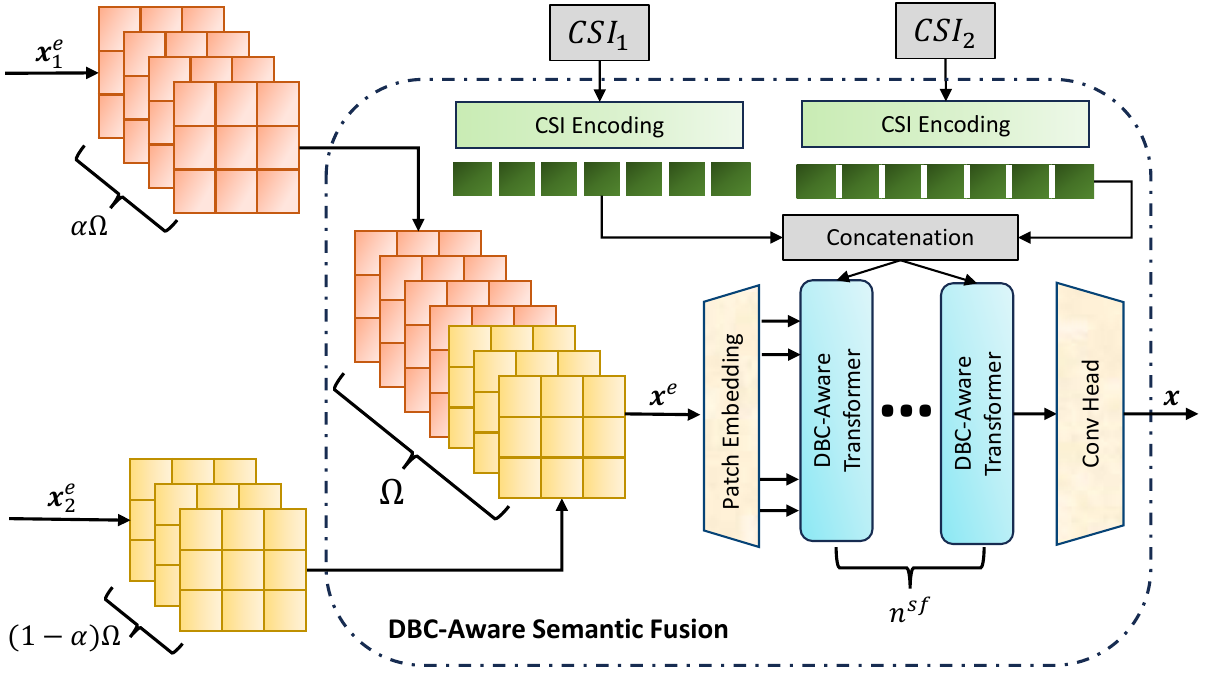}
    \caption{The architecture of DBC-Aware Semantic Fusion module}
    \label{SF}
\end{figure}
The fusion process consists of three stages. As shown in Fig. \ref{SF}, the DBC-Aware semantic fusion module initially concatenates two semantic features $\mathbf{x}_1^e$ and $\mathbf{x}_2^e$ as $\mathbf{x}^e \in \mathbb{R}^{\Omega \times h \times w}$, serving as the input for the patch embedding module. The patch embedding module utilizes a CNN network to perform a first-stage fusion of the semantic features from two users, jointly mapping them onto a high-dimensional feature map. In the second-stage fusion, $n_{sf}$ DBC-Aware Transformers use boths $CSI_1$ and $CSI_2$ for channel-adaptive fusion.
In the third stage, a CNN network fuses the semantic features as $\mathbf{x}$ in a compression manner for transmission over the DBC. It is essential that we employ the fusion ratio $\alpha$ to control the quantity of semantic features from two users entering the semantic fusion module, aiming to balance the reconstruction quality between the two users. Specifically, the total number of input channels in the semantic fusion module is $\Omega$, where $\lceil \alpha\Omega \rceil$ channels are attributed to the semantic features from User 1, meaning $c_1=\lceil \alpha \Omega \rceil$, while $\lfloor (1-\alpha)\Omega \rfloor$ channels consist of the semantic features from User 2, indicating $c_2=\lfloor (1-\alpha) \Omega \rfloor$. For the semantic de-fusion module at user $j$, it has the same structure as the first stage in DBC-Aware JSCC encoder, with $n^{df}$ DBC-Aware Transformers but only encoding $CSI_j$ with the CSI encoding module.

\subsection{Loss Function Design}
In the semantic communications system for image transmission over DBC, the better user expects to reconstruct its individual image $\mathbf{s}_1$ as much as possible. 
Thereby the network $f_e(\cdot\theta_1)$, $f_{sf}(\cdot;\varphi )$, $g_{df}(\cdot;\psi_1)$ and $g_d(\cdot;\phi_1)$ are jointly trained with the channel to minimize the distortion between $\mathbf{s}_1$ and $\hat{\mathbf{s}}_1$. The following loss function is derived as
\begin{equation}
    L_1(\theta_1,\varphi ,\psi _1,\phi _1)=\mathbb{E}_{\mathbf{s}_1 \sim p_s,\mathbf{y}_1 \sim p_{\mathbf{y}_1|\mathbf{x}}}\ d(\mathbf{s}_1,\hat{\mathbf{s}}_1).
\end{equation}
where $d(\cdot)$ denoted the distortion function.

Simlarly, the worse user wishes to reconstruct its image $\mathbf{s}_2$ with mininal distortion between $\mathbf{s}_2$ and $\hat{\mathbf{s}}_2$.  $f_e(\cdot\theta_2)$, $f_{sf}(\cdot;\varphi )$, $g_{df}(\cdot;\psi_2)$ and $g_d(\cdot;\phi_2)$ are jointly trained with the channel with loss function
\begin{equation}
    L_2(\theta_2,\varphi ,\psi _2,\phi _2)=\mathbb{E}_{\mathbf{s}_2 \sim p_s,\mathbf{y}_2 \sim p_{\mathbf{y}_2|\mathbf{x}}}\ d(\mathbf{s}_2,\hat{\mathbf{s}}_2)
\end{equation}

In this paper, we adopt PSNR as the semantic performance matric and thus the distortion function is MSE function. The relationship between them is:
\begin{equation}
    PSNR=10\log_{10}{\frac{1}{MSE}}.
\end{equation}

It can be observed that the module $f_{sf}(\cdot;\varphi)$ is optimized by both loss functions, indicating that a combination of the two loss function is necessary for network training. A classical combination approach involves introducing a weight control factor $\lambda$ and summing the two loss functions. However $\lambda$ is required to be selected carefully according to the fusion ratio $\alpha$ to achieve the largest semantic performance region.

Aiming to enlarge the semantic performance region, we combine the two loss function as $L_3$ as follows

\begin{align}
    L_3&(\theta_1,\phi _1,\psi _1,\theta_2,\phi _2,\psi _2,\varphi )=-(\log L_1)^2-(\log L_2)^2\nonumber\\
    &=-(\log \frac{1}{\mathbb{E} _{\mathbf{s}_1 \sim p_s,\mathbf{y}_1 \sim p_{\mathbf{y}_1|\mathbf{x}}}d(\mathbf{s}_1,\hat{\mathbf{s}}_1)}) ^2\nonumber\\
    &-(\log \frac{1}{\mathbb{E}_{\mathbf{s}_2 \sim p_s,\mathbf{y}_2 \sim p_{\mathbf{y}_2|\mathbf{x}}}d(\mathbf{s}_2,\hat{\mathbf{s}}_2)})^2.
\end{align}

This loss function implies that the objective of training the neural network is to move the semantic performance point, measured by PSNR, further away from the zero point in the semantic performance region. Moreover, the gradient of $L_3$ during training process is derived as
\begin{align}
     \nabla_{\theta_1,\phi _1,\psi _1,\theta_2,\phi _2,\psi _2,\varphi } L_3= &-\frac{2}{\ln10 }\frac{\log L_1}{L_1}\nabla_{\theta_1,\phi _1,\psi _1,\varphi } L_1\nonumber\\
     &-\frac{2}{\ln10}\frac{\log L_2}{L_2}\nabla_{\theta_2,\phi _2,\psi _2,\varphi } L_2.
\end{align}

\begin{figure*}[t]
    \centering
    \includegraphics[width=0.98\textwidth]{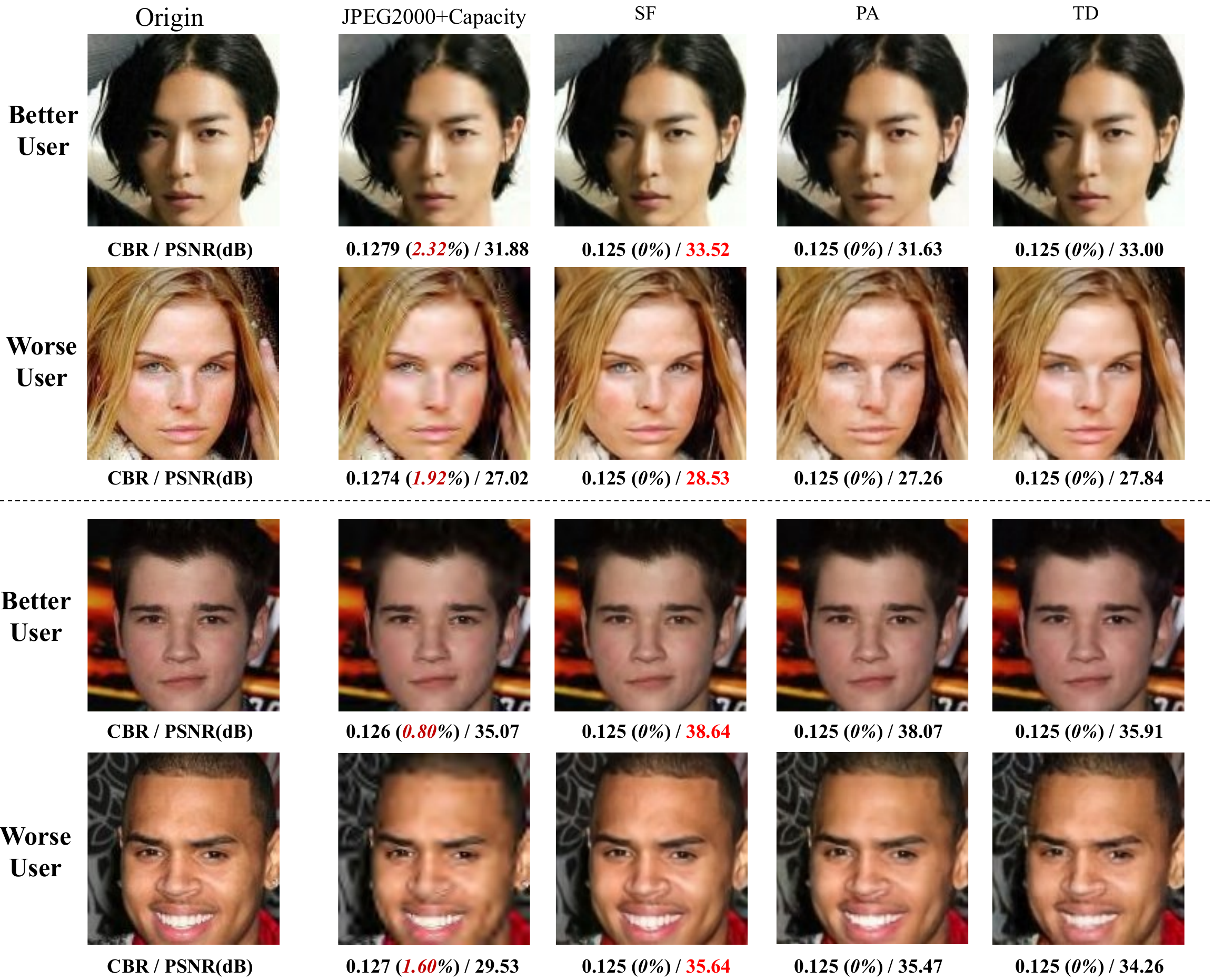}
    \caption{Examples of visualization results under DBC at SNR=$13$ dB for the better user and $8$ dB for the worse user. The five columns display the original images and the reconstructed images obtained from their respective schemes. The dark red number within parentheses corresponds to the percentage of additional bandwidth cost compared to the semantic communication schemes, and the light red PSNR value signifies the best performance of this image among all schemes.}
    \label{vis}
\end{figure*}

This suggests that, in contrast to utilizing a fixed hyperparameter $\lambda$ to regulate the weights of the two distortions, {this loss function enables the network to dynamically adjust the weights of the gradients of the two distortion functions based on their values during training, although it not directly measures the semantic performance region}. Thereby, the loss function $L_3$ is suitable for achieving the goal of maximizing the semantic performance region.

\section{EXPERIMENTS RESULTS}
In this section, we present experimental results to demonstrate the effectiveness of our proposed SF broadcasting scheme. Additionally, we assess the additional parameters and computational overhead introduced by various channel adaptive methods, thereby demonstrating the cost-effectiveness of our DBC-Aware approach in terms of deployment and inference.
\subsection{Experimental Setup}
\textbf{Dataset:} To obtain universally experimental results, {we adopt three image datasets with different resolutions. For low-resolutions images, we use CIFAR10 and STL-10 datasets \cite{CIFAR10,STL-10} with resized dimensions $32 \times 32$, including $50000$ color images for training and $10000$ for testing.} The high-resolution image dateset is CelebA dataset \cite{CelebA}, comprising of $150000$ color images about human faces for training and $50000$ images for testing. We crop the human-face images into a size of $128 \times 128$.

\textbf{Comparsion broadcasting schemes:} Due to the lack of in-depth research on semantic communications over DBC, we consider incorporating two traditional channel dividing schemes into semantic communications as benchmarks, namely TD and {power allocation (PA)}. For the TD scheme, the number of channel uses, denoted as $n$, is divided between the two users according to the TD allocation ratio $\beta$. Each user then transmits their individual signals, which are produced by their DBC-Aware JSCC encoders, over the assigned time slots. For the PA scheme, the total transmission power is divided between the two users according to the PA allocation ratio $\gamma $. Each user utilizes their own DBC-Aware JSCC encoder to generate their respective transmission signals. Subsequently, the transmitter broadcasts the two signals simultaneously with the corresponding power. The DBC-Aware JSCC encoders in both comparison schemes have the same structure as those in the proposed SF broadcasting scheme and decoders have the corresponding reverse structure. Both comparison schemes are jointly trained with the two users and the transmitter by combining the loss functions of two users using $\lambda=0.3$. Moreover, we also conduct a comparative analysis with the superposition coding with SIC broadcasting scheme with the power allocation coefficient $\zeta$. This broadcasting scheme is a capacity achieving scheme and we adopt JPEG2000 for source coding and capacity-achieving code for channel coding thus the scheme is denoted as ``JPEG2000+Capacity". 
\begin{figure}[t]
    \centering
    \includegraphics[width=0.48\textwidth]{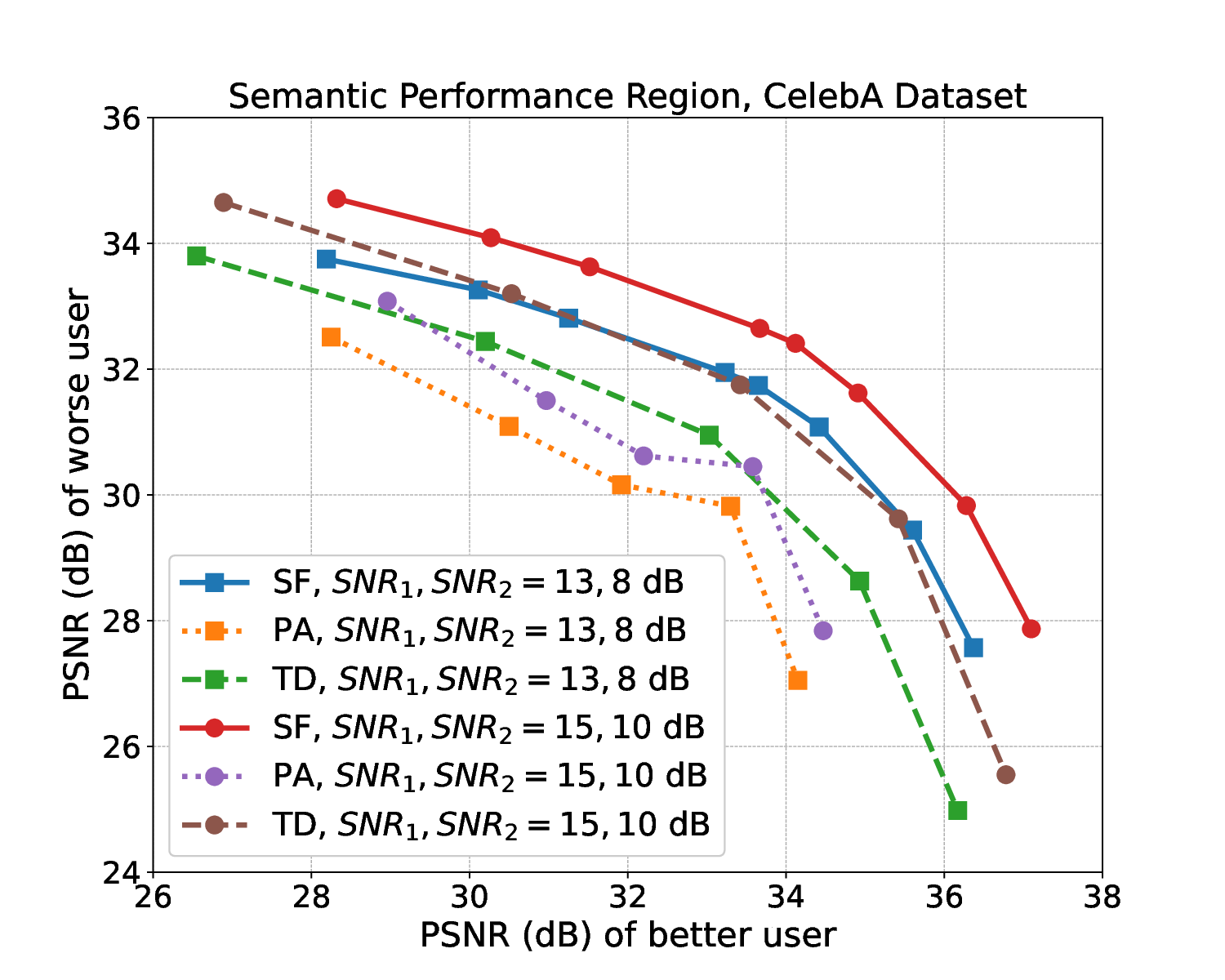}
    \caption{Semantic performance region of CelebA dataset with different broadcasting schemes.}
    \label{CelebA_region}
\end{figure}

\textbf{Comparsion channel adaptive method}: We compare our DBC-Aware channel adaptive method with the attention-based method proposed in \cite{KeYang}. We employ the two channel adaptive methods in the SF scheme for broadcasting, employing the same training strategy for both. Specifically, in the attention-based method, two channel attention modules are applied after the two JSCC encoders. Then, the transmitter integrates their outputs with SF modules for broadcasting. The two users de-fuse their received signals and employ two attention modules before their JSCC decoders, respectively. The JSCC encoders, decoders, the SF module and the de-fusion modules have the same structures as those in the proposed system, except for components related to CSI.

\textbf{Training details:} The configurations of the DBC-Aware JSCC, SF and de-fusion modules vary with training image resolution. For low-resolution images, we use one stage with $n_1^e=n_1^d=4$ for JSCC and $n^{sf}=n^{df}=2$ for SF and de-fusion modules. The channel bandwidth ratio (CBR) is set to $0.25$. {For high-resolution images, the semantic features are more complex than those of low-resolution images and thus required more complex model structure. As a result, we employ two stages with $n_1^e,n_2^e=2,4$ and $n_1^d,n_2^d=4,2$ for JSCC and $n^{sf}=n^{df}=6$ for SF and de-fusion modules.} The CBR is set to $0.125$. For AWGN channels, $CSI_1$ refers to the SNR of the better user, denoted as $SNR_1$, while $CSI_2$ refers to $SNR_2$. All the systems are trained across a range of SNRs from $5$ dB to $19$ dB for the better user and the SNR of the worse user is $3$ dB or $5$ dB worse than that of the better user, indicating that for each scheme, we evaluate its performance at different SNRs with same parameters.
We employ Adam optimizer \cite{adam} with a learning rate $0.0001$ to optimize all the systems.

\subsection{Visualization Results}
Fig. \ref{vis} visualizes two sets of reconstructed images generated by the four broadcasting schemes. The results are obtained under DBC with $\zeta=0.2$, $\alpha=\beta=\gamma=0.5$ and $SNR_1=13$ dB, while $SNR_2=8$ dB. We can observe that the SF broadcasting scheme achieves the best reconstruction quality, exhibiting visually superior clarity compared to the JPEG+Capacity scheme and the PA scheme, particularly for images of the worse user, despite the slightly higher channel resource usage of the JPEG+Capacity scheme. In the case of images reconstructed using the TD scheme, the images of the better user are visually nearly identical to those reconstructed using SF scheme. However, the images of the worse user lack many details. For instance, the image of the worse user in the first set misses details in the cheek region, while in the second set, the details of the skin around the corners of the mouth are lacking.
\subsection{Performance Analysis}
\begin{figure}[t]
    \centering
    \includegraphics[width=0.48\textwidth]{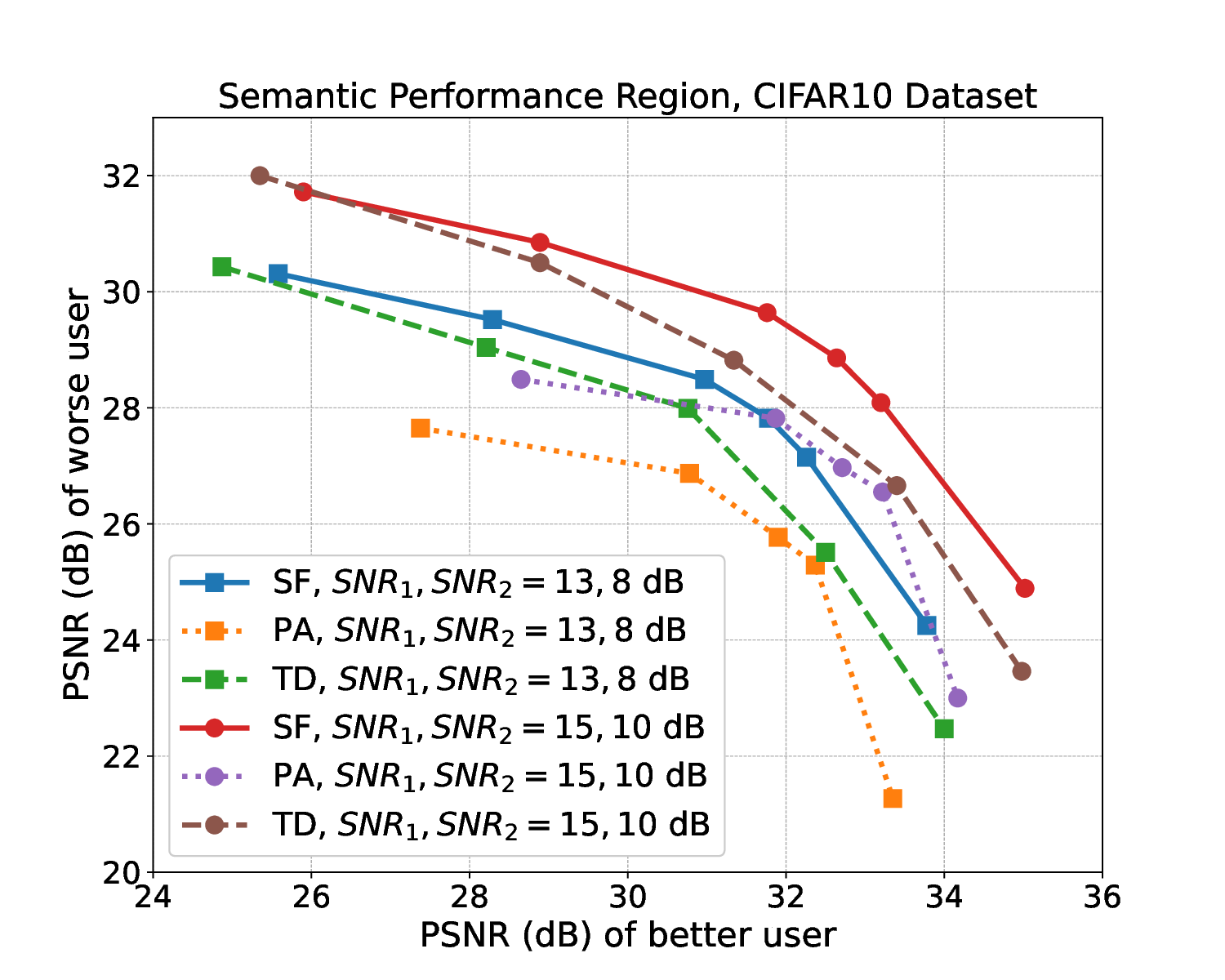}
    \caption{Semantic performance region of CIFAR10 dataset with different broadcasting schemes.}
    \label{CIFAR10_region}
\end{figure}
 Fig. \ref{CelebA_region} and Fig. \ref{CIFAR10_region} show the semantic performance regions characterized by PSNR based on CelebA and CIFAR10 datasets, respectively. The results marked with squares corresponds to $SNR_1=13$ dB and $SNR_2=8$ dB, while the dots represent results at $SNR_1=15$ dB and $SNR_2=10$ dB. For the SF scheme, adjusting $\alpha$ allows us to balance the PSNR performance between the two users. Similarly, for the the TD or PA schemes, this balance can be achieved by adjusting $\beta$ or $\gamma$. For example, in the SF scheme, increaseing $\alpha$ improves the PSNR performance of the better user, but decreases it for the worse user. 

It is evident from Fig. \ref{CelebA_region} and Fig. \ref{CIFAR10_region} that the semantic performance regions of the SF scheme encompass those of the TD and PA schemes under both SNR conditions and datasets. This demonstrates the effectiveness of the proposed SF scheme in fusing semantic features from two users for broadcasting by leveraging their semantic similarity, as illustrated in Fig. \ref{RV2_coeff}. Thus, the SF scheme achieves the best performance for both users and both datasets in DBC compared with banchmarks. 
\begin{figure}[t]
    \centering
    \includegraphics[width=0.435\textwidth]{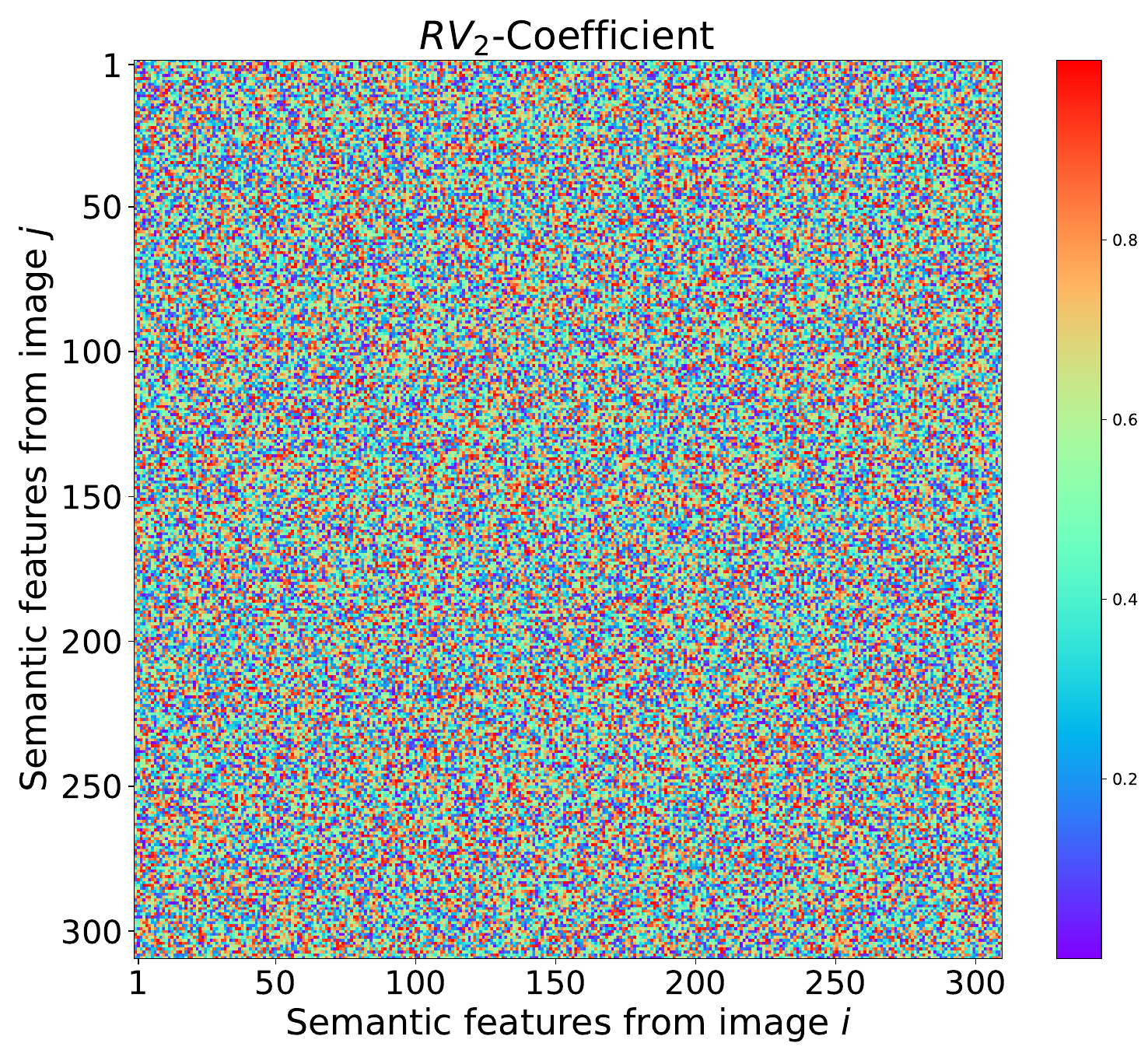}
    \caption{The $RV_2$-coefficient of semantic features from images in CIFAR10 and STL-10 datasets.}
    \label{rv2_merge}
\end{figure}

\begin{figure}[t]
    \centering
    \includegraphics[width=0.48\textwidth]{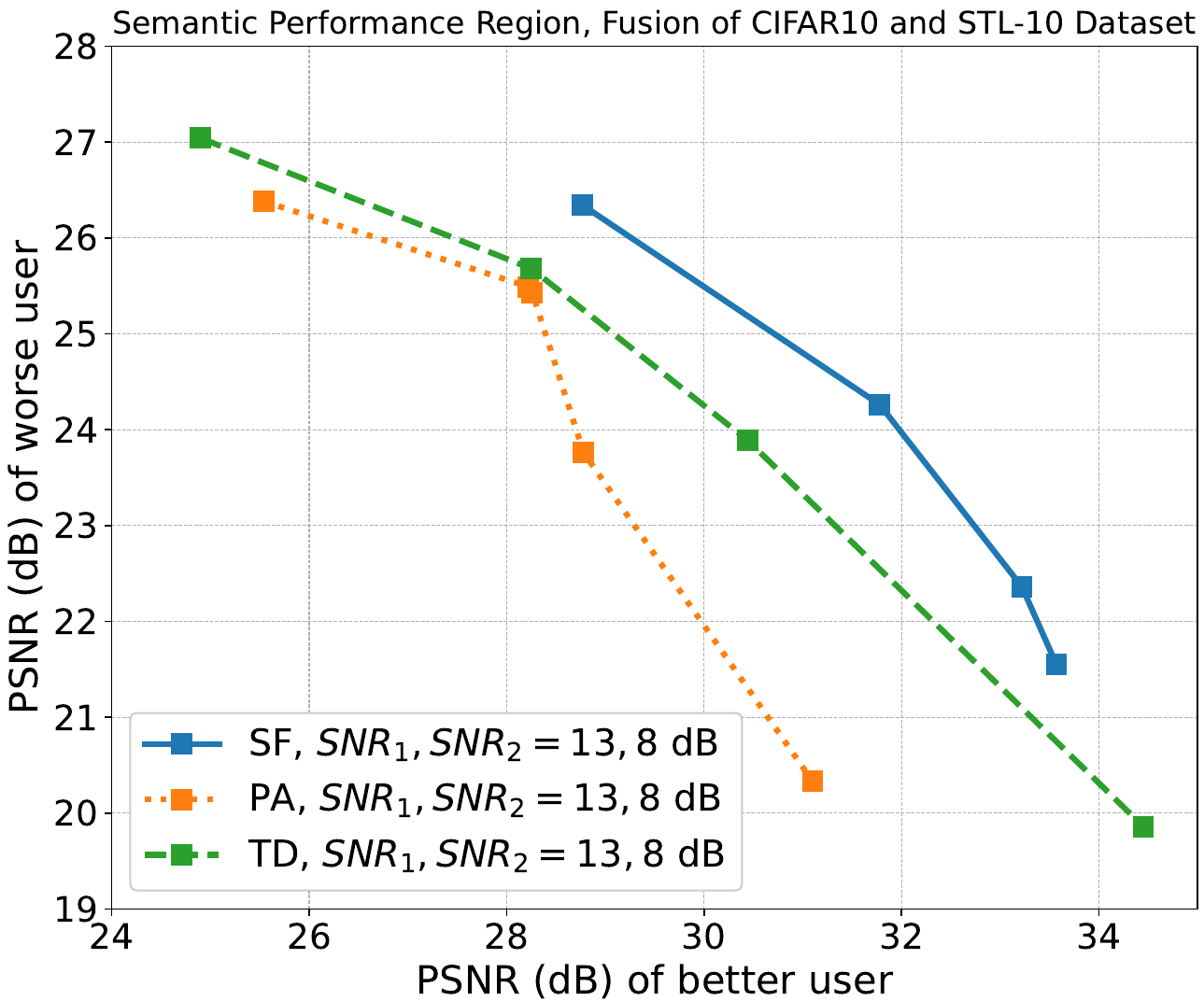}
    \caption{Semantic performance region with different broadcasting schemes. The dataset for better user is CIFAR10 and for worse user is STL-10.}
    \label{merge_region}
\end{figure}
{Furtherly, to extend our SF scheme to scenarios with lower semantic similarity in source images, we employ CIFAR10 for the better user and STL-10 for the worse user. Fig.\ref{rv2_merge} shows the $RV_2$ coefficients of the semantic features from the two datasets and we can discover that the mutual semantic similarity is lower than CelebA datasets bacause the two different datasets consist of diversity of images with different distributions. 
Fig.\ref{merge_region} illustrates the semantic performance regions of different broadcasting schemes with CIFAR10 for the better user and STL-10 for the worse user. The SNR is fixed at 13 dB for better user and 8 dB for the worse user. It can be observed that the SF scheme still achieves the largest semantic performance region among the three schemes, proving that out SF scheme is not limited in the scenarios with high semantic similarity and can be generally adopted in broadcasting scenarios.}
\begin{figure}[t]
    \centering
    \includegraphics[width=0.48\textwidth]{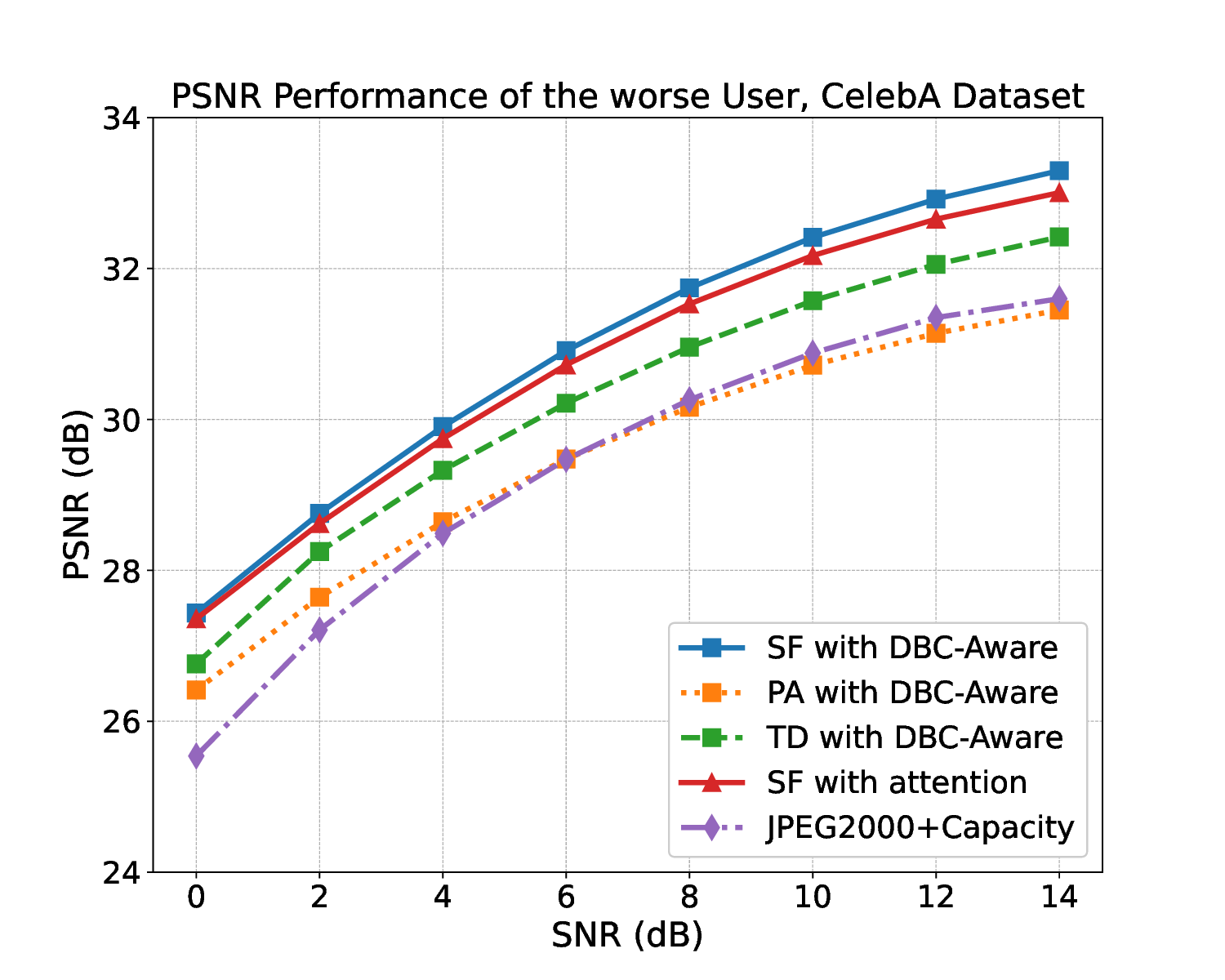}
    \caption{PSNR performance of the worse user versus SNR using CelebA dataset.}
    \label{CelebA_PSNR_worse}
\end{figure}
Moreover, the PA scheme generally shows the poorest performance, despite one power allocation ratio showing slightly improved performance compared to the others in CelebA dataset. This can be attributed that PA scheme combines them linearly for transmissions. However, as Fig. \ref{Linear_coeff} shows, the linear correlation between the semantic features from the two users is relatively low.

\begin{figure}[t]
    \centering
    \includegraphics[width=0.887\linewidth]{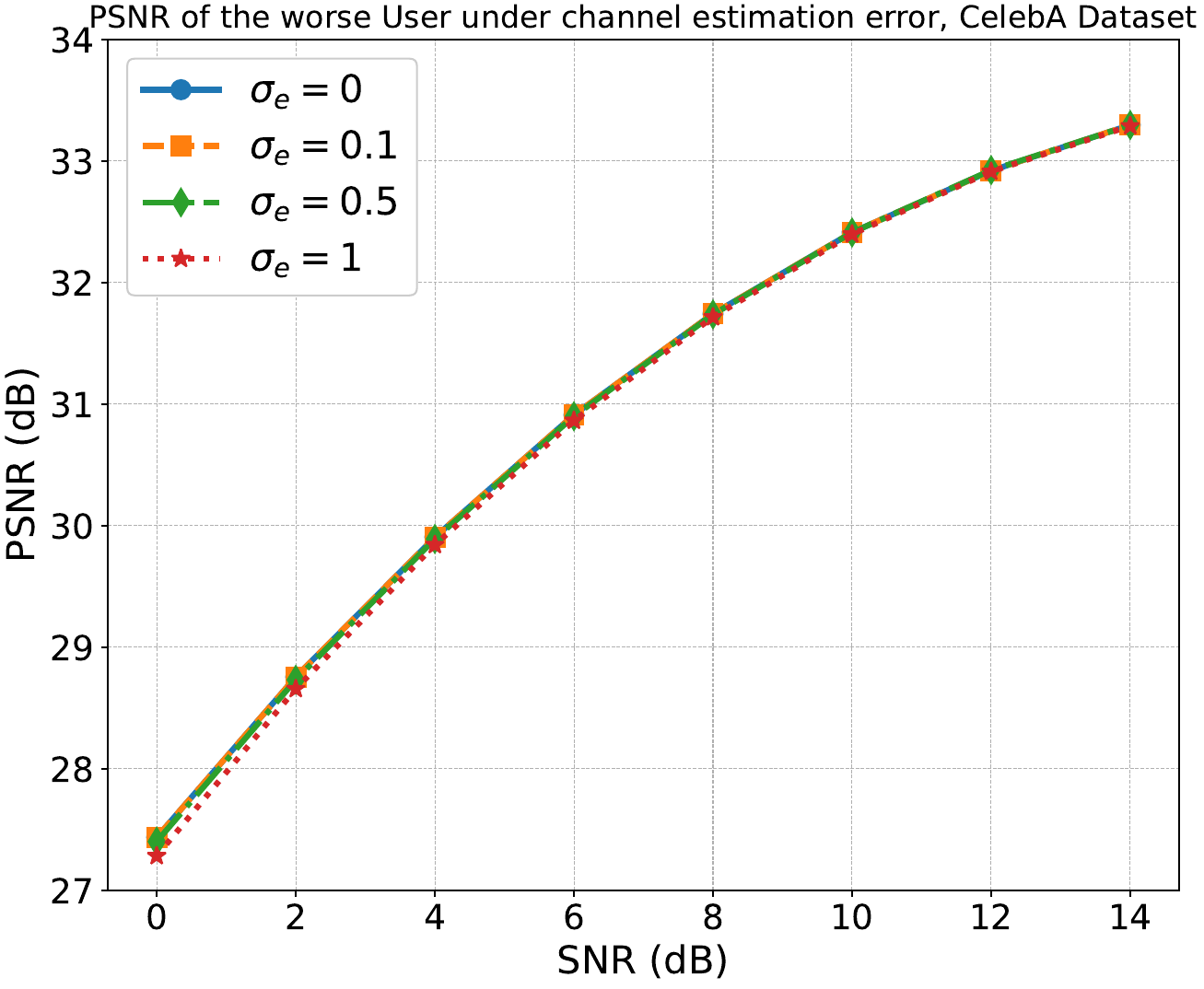}
    \caption{{PSNR of the worse user with channel estimation errors.}}
    \label{error_worse}
\end{figure}
\begin{figure}[t]
    \centering
    \includegraphics[width=0.48\textwidth]{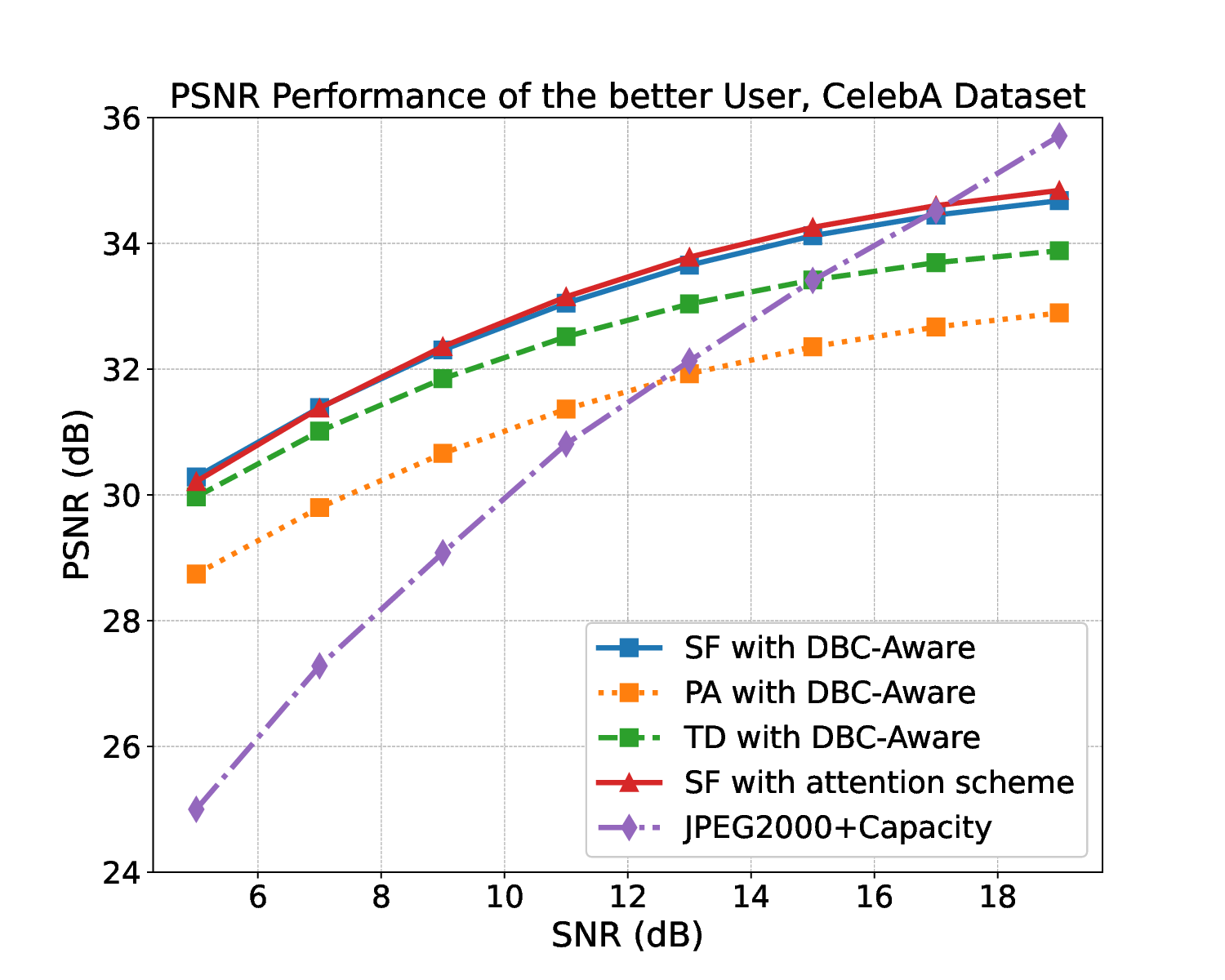}
    \caption{{PSNR performance of the better user versus SNR using CelebA dataset.}}
    \label{CelebA_PSNR_better}
\end{figure}

Fig. \ref{CelebA_PSNR_worse} and Fig. \ref{CelebA_PSNR_better} illustrate the PSNR performance of the better user and worse user, respectively, versus SNR under different broadcasting schemes and channel adaptive methods in CelebA dataset. The allocation ratios are set as $\alpha=\beta=\gamma=0.5$ and $\zeta=0.2$. $SNR_2$ is $5$ dB lower than $SNR_1$. It can be observed that with the same DBC-Aware channel adaptive method, the SF scheme outperforms the TD and PA schemes in terms of PSNR performance. Moreover, the performance of both users with JPEG2000+Capacity scheme is lower than that with the SF scheme in the low SNR range. For high SNR, e.g., $SNR_1=19$ dB and $SNR_2=14$ dB, the performance of JPEG2000+Capacity scheme exceeds that of the SF scheme for the better user by $1.04$ dB. However, it still exhibits a performance gap of $1.69$ dB for the worse user.

We can also see from Fig.\ref{CelebA_PSNR_worse} and Fig.\ref{CelebA_PSNR_better} that in the SF scheme, the PSNR performance of the worse user based on the DBC-Aware method slightly exceeds that of the attention-based approach. This difference is particularly noticeable at higher SNRs, e.g., a gain of $0.30$ dB at $SNR_1=14$ dB. Conversely, for the better user, the performance achieved by both the DBC-Aware method and the attention-based method is almost the same. The largest observed difference is minimal, e.g., at just $0.14$ dB at $SNR_2=19$ dB. In summary, the DBC-Aware method slightly outperforms the attention-based method in terms of the PSNR performance for both users on the CelebA dataset. 
Furthermore, Table \ref{tab1} presents the multiply accumulates (MACs) operations, {interference delay (ID)} and parameters of the models employing the SF scheme under different channel adaptive methods. {The MACs and parameters are calculated by Pytorch-OpCounter. The ID is obatined on a single NVIDIA A40 GPU and only includes the time cost of the inference process of the transmitter and two receivers. On the CelebA dataset, the DBC-Aware method introduces an addition of $6$M MACs, $0.59$ms ID and $1.31$M parameters compared to the approach without CSI. In constrast, the attention-based method requires an addition of $1300$M MACs, $4.34$ms ID and $16.18$M parameters. This indicates that the proposed DBC-Aware method not only achieves better performance but also significantly computational and parameter efficiency}.

\begin{figure}[t]
    \centering
		\includegraphics[width=0.9\linewidth]{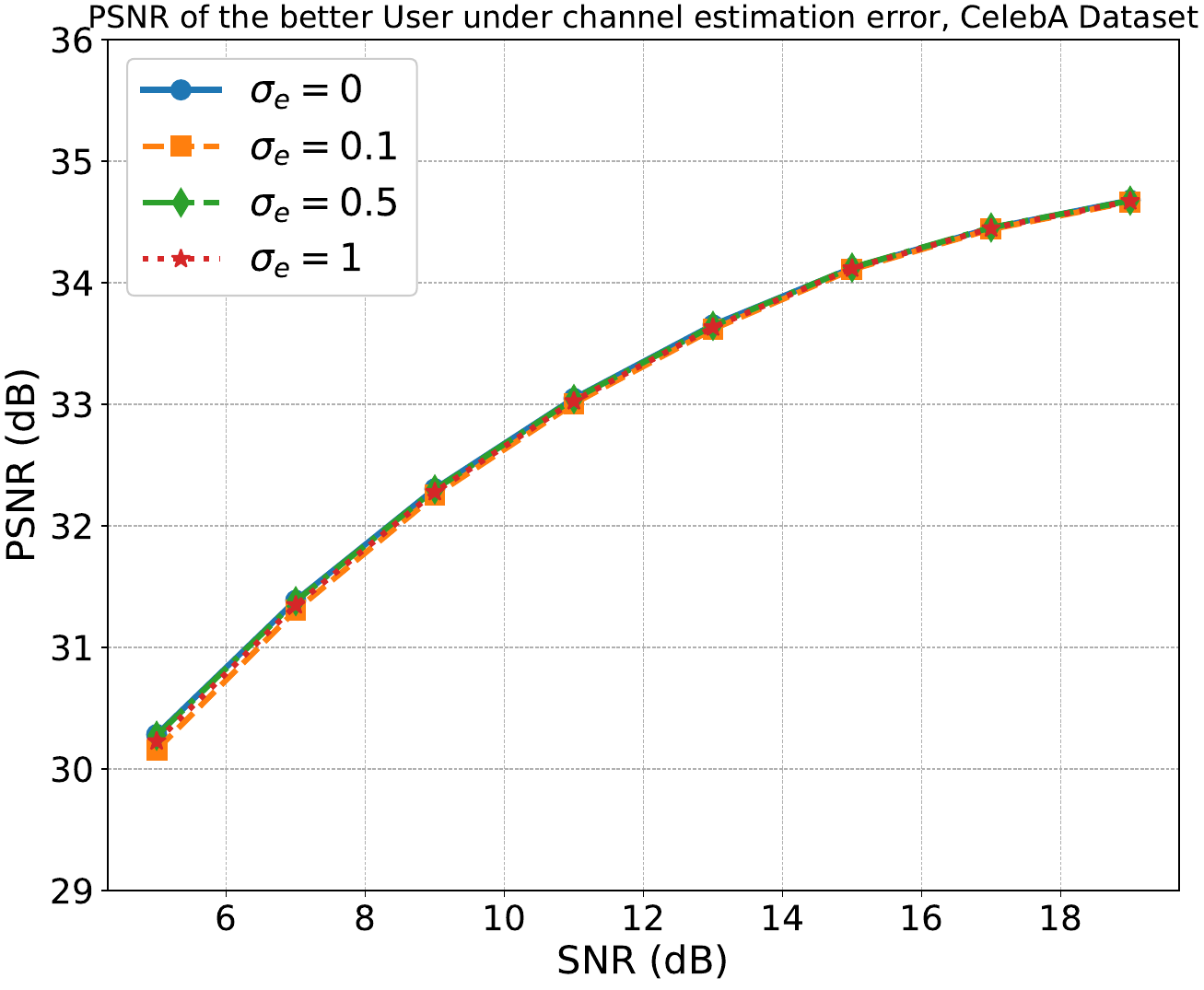}
		\caption{PSNR of the better user with channel estimation errors.}
		\label{error_better}
\end{figure}
\begin{table*}[t]
    \centering
    \caption{{MACs, Interference Delay, Parameters of the fusion-based systems employed on two datasets with different channel adaptive methods.}}
    \label{tab1}
    \begin{tabular}{|c|c|c|c|c|c|c|}\hline
    
    \  & \multicolumn{3}{c|}{CelebA} & \multicolumn{3}{c|}{CIFAR10}\\
    \hline
    \  & MACs &ID&Parameters & MACs &ID &Parameters\\\hline
    No CSI & 25.906G &1.53 ms&39.303M & 698.6M &0.189 ms& 4.105M\\
    DBC-Aware & 25.912G &2.12 ms&40.608M & 699.2M &0.227 ms& 4.335M\\
    Attention & 28.996G &5.87 ms&55.486M & 873.7M &0.440 ms& 7.278M\\\hline
    \end{tabular}
\end{table*}
{To furtherly investigate the performance of our DBC-Aware method under channel estimation errors and different SNR gap, we provide corresponding experimental results in Fig .\ref{error_worse}, \ref{error_better}-\ref{gap_better}. More specifically, in the scenario where channel estimation errors exist, we set the SNR changing randomly, which can be formulated as:
\begin{align}
    SNR_1=\hat{SNR_1}+\Delta SNR_1\nonumber\\
    SNR_2=\hat{SNR_2}+\Delta SNR_2,
\end{align}
\begin{figure}[ht]
    \centering
    \includegraphics[width=0.9\linewidth]{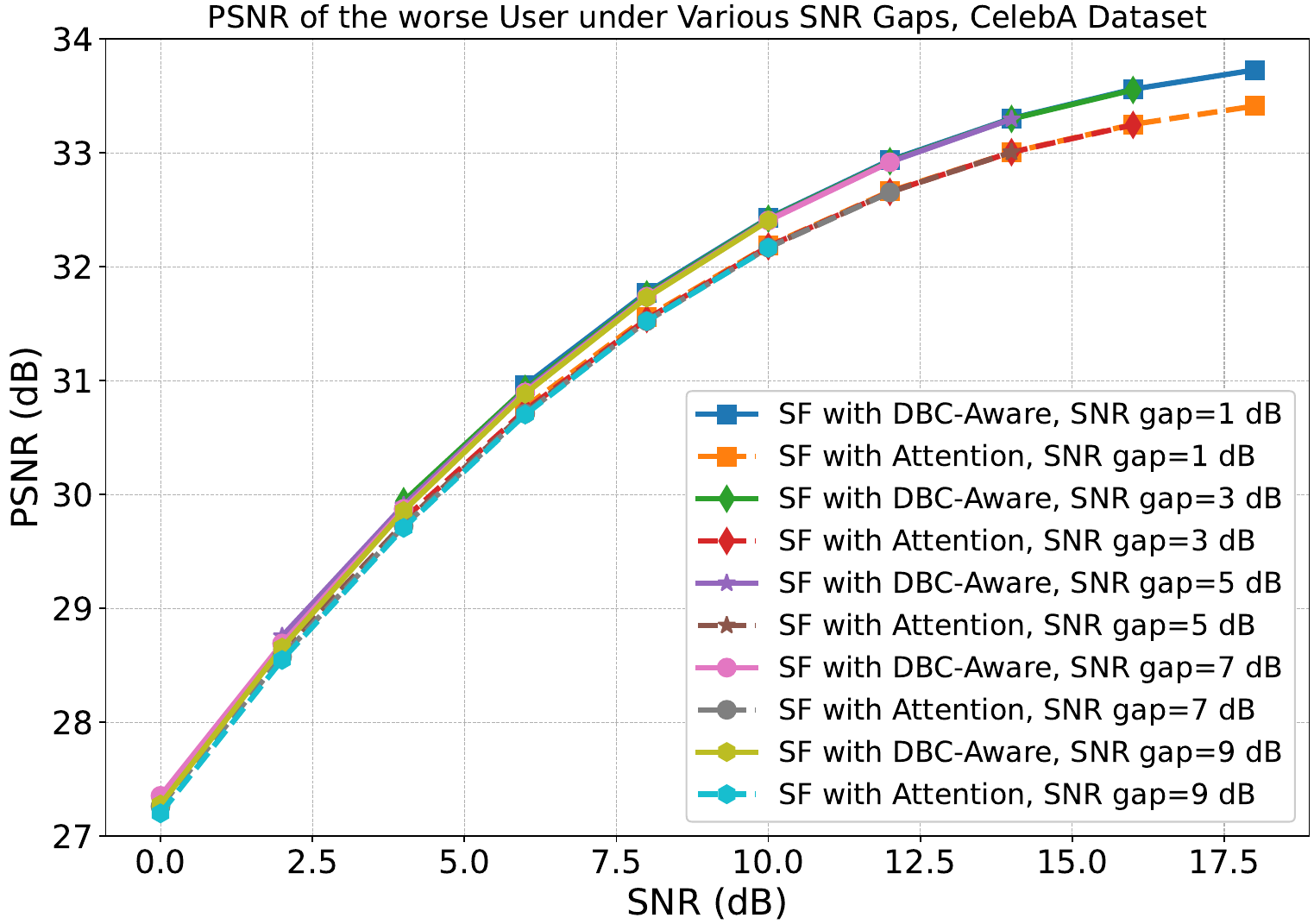}
    \caption{{PSNR of the worse user with varying difference in SNRs.}}
    \label{gap_worse}
\end{figure}
where $\hat{SNR1}$ and $\hat{SNR2}$ are the channel estimation results and $\Delta SNR1$, $\Delta SNR2 \sim N(0, \sigma_e^2)$ reflect the channel estimation errors. Fig.\ref{error_worse} and \ref{error_better} show the PSNRs of the worse user and better user with different $\sigma_e$ in CelebA dataset, where the difference of the two SNRs is fixed at 5 dB. It can be discovered that as the increase in $\sigma_e$, only slight decrease in PSNR can be observed. For example, when the $\sigma_e$ increase from $0$ to $1$, the PSNR of the better/worse user decreases from $30.28$/$27.43$ dB to $30.15$/$27.12$ dB at most, which is observed at $\hat{SNR_1}=5$ dB and $\hat{SNR_2}=0$ dB. While at minimum, the PSNR of the better/worse user only decreases by $0.014$/$0.018$ dB when $\hat{SNR_1}$ is 19 dB and $\hat{SNR_2}$ is 14 dB. The results demonstrate that our system maintains good performance in the presence of channel estimation errors. 
}

{Furtherly, we consider changing the SNR gaps, as shown in Fig.\ref{gap_worse} and \ref{gap_better}, where the difference in SNR is varying from $1$ dB to $9$ dB.
It can be observed that generally, our model maintains almost consistent performance to the changing in the difference of SNRs. For specific,
\begin{figure}[ht]
    \centering
		\centering
		\includegraphics[width=0.9\linewidth]{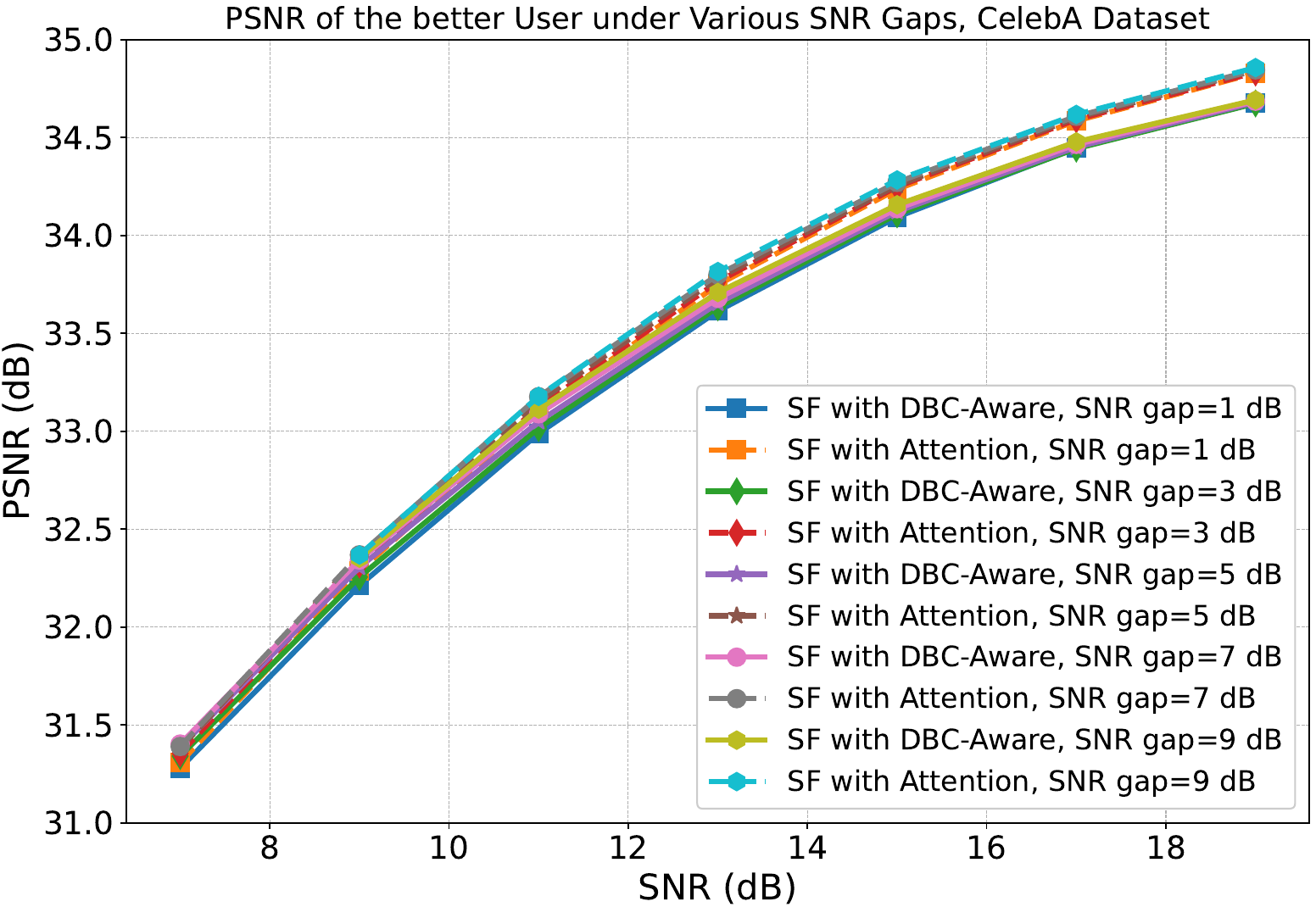}
		\caption{{PSNR of the better user with varying difference in SNRs.}}
		\label{gap_better}
\end{figure}
when $SNR_1$ is fixed and the $SNR_2$ decrease, meaning the difference is increase, the performance of user 1 increase and the maximum increase of $0.14$ dB occurs when $SNR_1$ is $9$ dB. At this condition, when $SNR_2$ is $1$ dB lower than $SNR_1$, the PSNR of user 1 is $32.21$ dB. While $SNR_2$ is $9$ dB lower than $SNR_1$, the PSNR of user 1 increase to $32.35$ dB. 
The experimental results illustrate that our system is robust to the changing in the differences of SNRs and a large difference is helpful for the de-fusion the semantic features, resulting in slightly better reconstruction quality.
}
{Additionally, compared with the performance of attention-based channel adaptive method, our method perform at most $0.15$ dB worse than the attention-based channel adaptive method for the better user but $0.19$ dB better for the worse user. Therefore, generally, our DBC-Aware channel adaptation method and the attention-based channel adaptation method achieve slight better performance, and as illustrate in Table \ref{tab1}, our DBC-Aware channel adaptation method is more efficient in computational complexity and the amount of parameters.}

\begin{figure}[t]
    \centering
    \includegraphics[width=0.48\textwidth]{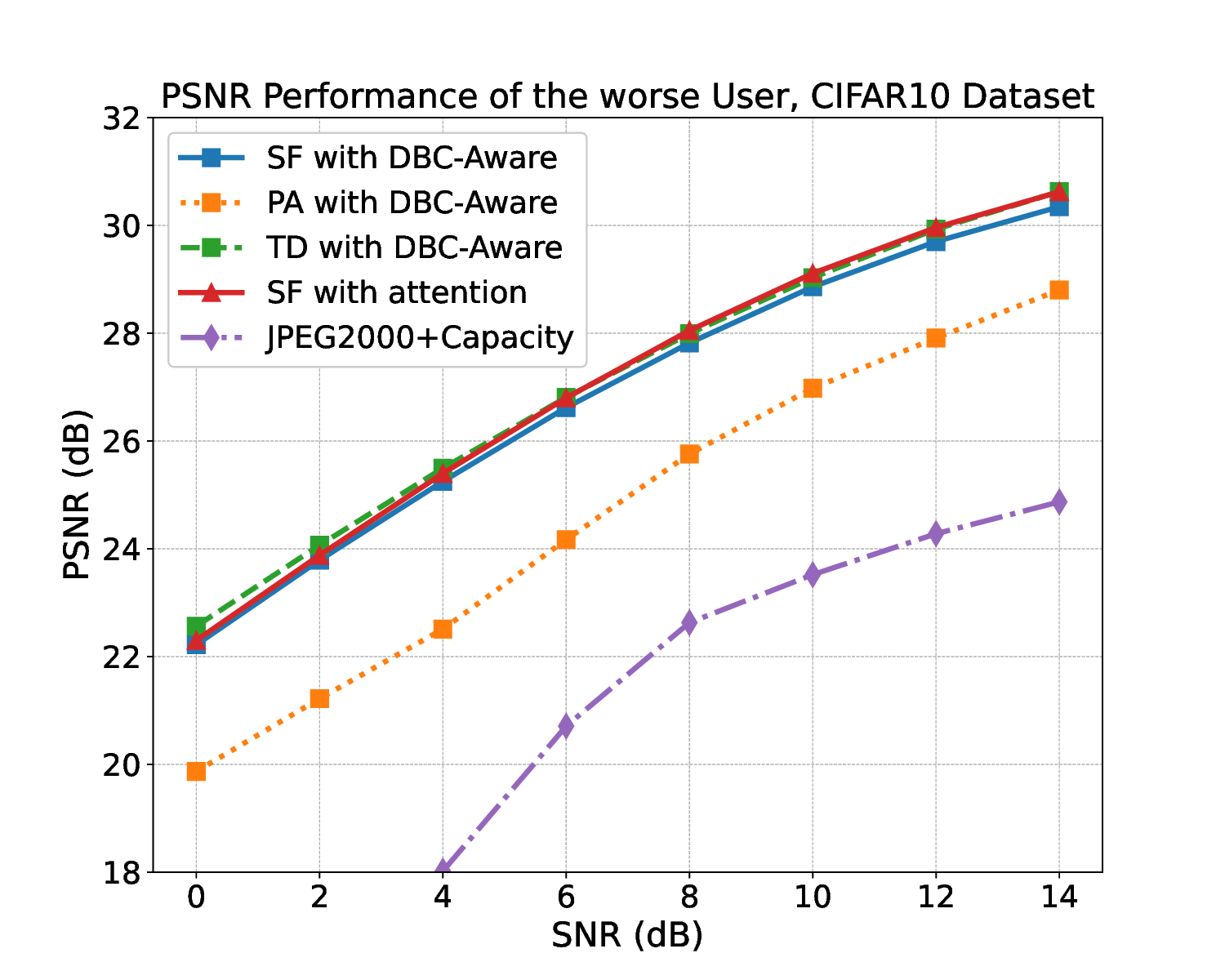}
    \caption{PSNR performance of the worse user versus SNR using CIFAR10 dataset.}
    \label{CIFAR10_PSNR_worse}
\end{figure}

\begin{figure}[t]
    \centering
    \includegraphics[width=0.48\textwidth]{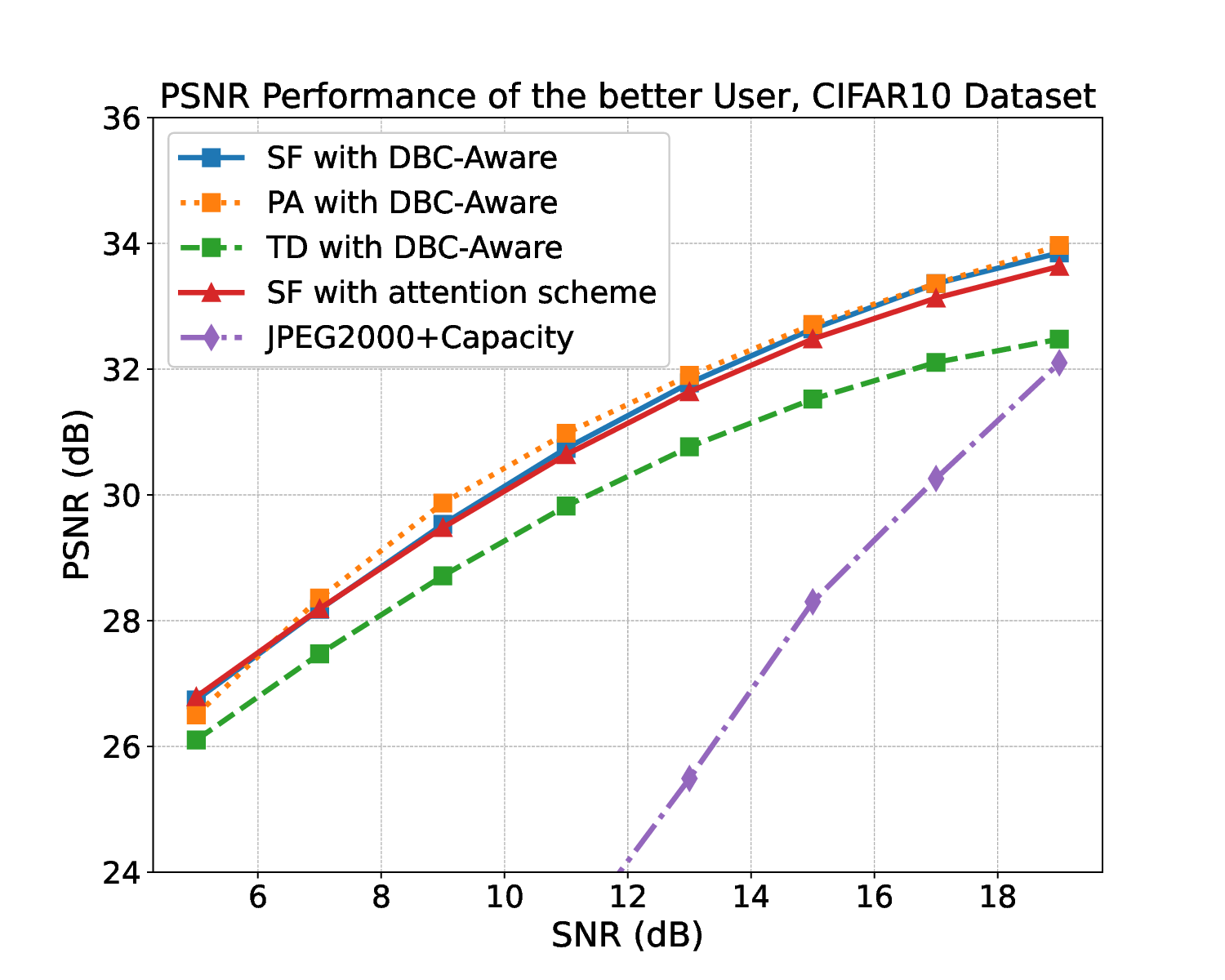}
    \caption{PSNR performance of the better user versus SNR using CIFAR10 dataset.}
    \label{CIFAR10_PSNR_better}
\end{figure}

Fig. \ref{CIFAR10_PSNR_worse} and Fig. \ref{CIFAR10_PSNR_better} depict the PSNR performance of the better and worse users, respectively, versus SNR under different broadcasting schemes and channel adaptive methods in CIFAR10 dataset. The settings for the allocation ratios are consistent with those presented in Fig. \ref{CelebA_PSNR_worse} and Fig. \ref{CelebA_PSNR_better}. It is observed that although the TD scheme shows similar PSNR performance to the SF scheme for the worse user, its significantly lags behind the SF scheme for the better user. Conversely, the PA scheme achieves slightly better PSNR performance than the SF scheme for the better user, but its performance for the worse user is significantly lower. This indicates that achieving comparable performance to the SF scheme for one user comes at the cost of reduced performance for the other user.

We can see that the JPEG2000+Capacity scheme shows the poorest performance for both users in the CIFAR10 dataset. In terms of channel adaptation, the DBC-Aware method shows slightly better performance than the attention-based method for the better user. However, the attention-based method performs slightly better than the DBC-Aware method for the worse user. Therefore, on the CIFAR10 dataset, their performances are nearly identical. However, as shown in Table \ref{tab1}, similar to the results on the CelebA dataset, the DBC-Aware method introduces only {an addition of $0.6$M MACs, $0.038$ms ID and $0.23$M parameters, while the attention-based method incurs an addition of $175$M MACs, $0.215$ms ID and $3.17$M parameters.} Therefore, the DBC-Aware method significantly reduces parameter and computational complexity compared with attention-based method.


{\section{CONCLUSION and FUTURE WORK}}
In this paper, we have proposed a novel multiuser semantic communication system for wireless image transmission over degraded broadcast channels. The transmitter is capable of extracting semantic features from both users and effective fusing them using our developed semantic fusion module. We have incorporated the proposed DBC-Aware channel adaptive method into the system, enabling stable performance across varing SNRs with a single model while incurring low computational and parameter overhead. Experimental results have shown that the proposed system significantly outperforms traditional transmission schemes such as TD and PA, achieving effective channel adaptation with few extra multiply accumulates and parameters. 
{In the future, we will explore the potential generalizability of our approach to multi-user broadcasting scenarios}

\bibliographystyle{gbt7714-numerical}
\bibliography{myref}

\end{document}